\documentclass[11pt]{article}

\usepackage{array} 
\usepackage{amssymb}
\usepackage{graphics,graphpap}
\usepackage{graphicx}
\usepackage{color}
\usepackage{graphicx}
\usepackage{dcolumn}
\usepackage{epsfig}
\usepackage{epstopdf}
\usepackage{bm}
\usepackage{amsmath}
\usepackage{amsfonts}
\usepackage{setspace}

\renewcommand{\thefootnote}{\fnsymbol{footnote}}

\newcommand{\bmp}{\noindent\begin{minipage}{16cm}}
\newcommand{\emp}{\end{minipage}\vskip 7mm} 

\def\drawbox#1#2{\hrule height#2pt
        \hbox{\vrule width#2pt height#1pt \kern#1pt
              \vrule width#2pt}
              \hrule height#2pt}

\def\Asym#1#2{\vcenter{\vbox{\drawbox{#1}{#2}
              \kern-#2pt 
              \drawbox{#1}{#2}}}}

\def\simge{\mathrel{%
   \rlap{\raise 0.511ex \hbox{$>$}}{\lower 0.511ex \hbox{$\sim$}}}}

\def\simle{\mathrel{
   \rlap{\raise 0.511ex \hbox{$<$}}{\lower 0.511ex \hbox{$\sim$}}}}

\def\s#1{\setbox0=\hbox{$#1$}%
\rlap{\ifdim\wd0>.7em\kern.22\wd0\else\kern.1\wd0\fi /}#1}

\newcommand{\TeV}{\,\mbox{TeV}}
\newcommand{\GeV}{\,\mbox{GeV}}

\newcommand{\z}[1]{\mathbb{Z}_{#1}}
\newcommand{\R}[1]{\text{\bf#1}}
\newcommand{\Rb}[1]{\bar{\text{\bf#1}}}
\newcommand{\Rbb}[1]{\overline{\text{\bf#1}}}
\newcommand{\Ri}[2]{{\text{\bf#1}}_{\text{\bf #2}}}
\newcommand{\Rib}[2]{\bar{{\text{\bf#1}}}_{\text{\bf #2}}}
\newcommand{\mi}{\!-\!}
\newcommand{\I}{{\cal I}}

\newcommand{\iso}{\thickapprox}
\newcommand{\yuk}[1]{Y_{#1}}
\newcommand{\tran}{\Delta}

\begin{document}

\begin{titlepage}
\begin{flushright}\begin{tabular}{l}
SHEP 09-16 \\ 
\end{tabular}
\end{flushright}

\vskip1.5cm
\begin{center}
   {\Large \bf \boldmath On discrete Minimal Flavour Violation} 
    \vskip1.3cm {
    \sc Roman Zwicky $^{a}$\footnote{R.Zwicky@soton.ac.uk}  and
    Thomas Fischbacher $^{b}$\footnote{T.Fischbacher@soton.ac.uk}  
}
  \vskip0.5cm
$^a$ {\sl School of Physics \& Astronomy} \\
$^b$ {\sl School of Engineering }  \\[0.1cm]
University of Southampton \\  
Highfield, Southampton SO17 1BJ, UK
\vspace*{1.5mm}
\end{center}

\vskip0.6cm

\begin{abstract}
We investigate the consequences of replacing the global flavour symmetry
of Minimal Flavour Violation (MFV) $SU(3)_{Q} \times SU(3)_{U} \times SU(3)_{D} \times ..$
by a discrete  
${\cal D}_{Q} \times {\cal D}_{U} \times {\cal D}_{D} \times ..$ symmetry. 
Goldstone bosons  resulting from the breaking of the flavour symmetry generically lead to 
bounds on new flavour structure many orders of magnitude above the TeV-scale. The absence of Goldstone bosons for
discrete symmetries constitute the \emph{primary} motivation of our work.
Less symmetry implies further invariants  and  renders the mass flavour basis transformation  observable in principle and calls for a hierarchy in the Yukawa matrix expansion.
We show, through the dimension of the representations, that the (discrete) symmetry  in principle  does allow 
for additional $\Delta F  = 2$ operators. 
If though the  $\Delta F  = 2$ transitions are generated
by two subsequent $\Delta F = 1$ processes, as for example in the Standard Model, then the four crystal-like groups
$\Sigma(168) \iso {\rm PSL}(2,\mathbb{F}_7)$, $\Sigma(72 \varphi)$, $\Sigma(216 \varphi)$ and especially 
$\Sigma(360 \varphi)$ do provide enough protection  for a TeV-scale discrete MFV scenario. Models where this is not
the case have to be investigated case by case.
Interestingly $\Sigma(216\varphi)$ has a (non-faithful) representation corresponding to an  
$A_4$-symmetry.
Moreover we argue that the, apparently often omitted, $(D)$-groups
are subgroups of an appropriate $\Delta(6g^2)$.
We would like to stress that we do not provide an actual model that realizes the 
MFV scenario nor any other theory of flavour. 
\end{abstract}

\nonumber

\end{titlepage}

\newpage

\tableofcontents

\setcounter{footnote}{0}
\renewcommand{\thefootnote}{\arabic{footnote}}

\newpage

\section{Introduction}
\label{sec:intro}


In the absence of Yukawa interaction, $G_F = U(3)^5 = U(3)_Q \times U(3)_{U_R} \times 
U(3)_{D_R} \times U(3)_L \times U(3)_{E_R}$ is the maximal global symmetry
that commutes with the gauge groups of the Standard Model (SM)
 \cite{georgi_chivukula}.  The Yukawa matrices $\yuk{U,D,E}$ break this symmetry down
 to\footnote{The further breaking of this group down to $U(1)_{B-L}$ due to the chiral anomaly \cite{tHooft} 
is not central to this work.} 
\begin{equation}
\label{eq:yuk_break}
G_F = U(3)^5 \stackrel{\yuk{U,D,E}}{\longrightarrow} U(1)_B \times U(1)_L \;.
\end{equation} 
It was realized a long time ago \cite{technigim} that these sort of flavour symmetries forbid
flavour changing neutral currents (FCNC) at tree level. Most models of new physics do not posses
a rigid flavour structure and large FCNC effects should be expected in general.
On the other hand experiments in the quark flavour sector CLEO, BaBar, Belle,  NA48, KTeV, 
KLOE, TeVatron, .. do not, yet, show any significant deviations
of FCNC or CP-violation. This motivated the  \emph{effective field theory} approach, 
called Minimal Flavour Violation (MFV) \cite{MFV},  where it is postulated that the
sole sources of flavour violation are the Yukawa matrices. We shall be more precise later on. 
New physics contributions of the MFV-type compared to the SM in 
$K_0$ and $B_d$ oscillations can in principle still be as large as 
$C_{\rm SM}/C_{\rm MFV} \simeq  (0.5  \TeV) ^2 / m_W^2$ 
\cite{MFV}\footnote{We have to keep in mind that by quoting the scale 
$0.5  \TeV$ we implicitly assume a loop suppression factor as in the SM.
Besides loop suppression factors the actual scale of new physics 
is further masked by renormalization group effects as well as possible mixing angles of the underlying theory. Somewhat stronger bounds can be found
in a more recent  investigation \cite{UTfit}.}. 
It is the size of the Wilson coefficient $C_{\rm MFV}$, including its implicit loop suppression factor,  what we refer to as ``TeV-scale MFV scenario''. 
The extension of the concept from the quark sector to the lepton sector depends on the nature 
of the neutrino masses \cite{LFV}. For notational
simplicity we shall focus in this work on the quark sector. The results can easily be transfered to the
lepton sector.

Even in the absence of the knowledge of the exact dynamics one delicate 
question might be raised: How is the $G_F$ symmetry broken?
If the symmetry is broken spontaneously at some scale $\Lambda_F$, then this  gives rise to $3 \cdot 8 + 2 = 26$  CP-odd massless Goldstone bosons\footnote{Bearing in mind mass contributions from 
explicit breaking and anomalous U(1) factors.}   associated with the breaking of  
$U(3)^3 \to U(1)_B$. In connection with non-abelian  
family symmetries such Goldstone bosons are known
as familons \cite{familons}; they are  formally similar to an
axion  arising  from  the breaking of an axial U(1)-symmetry. Due to the fact that they are Goldstone
bosons their interactions with SM fields take on the universal form
${\cal L}_{\rm eff} \sim  \frac{1}{\Lambda_F}  \bar D_L T^a \gamma_\mu D_L  \partial^\mu\phi_F^a $, 
where $\phi_F^a$ is a familon and $T^a$ is the generator of the broken global symmetry.
The breaking scale\footnote{Related to the familon 
decay constant $f_F$ as follows: $\Lambda_F \simeq 4 \pi f_F$ \cite{Manohar:1983md}.}$\Lambda_{F}$  
is subjected to experimental constraints. Focusing on $s \to d$ transitions, which would give rise to the lowest scale in a scenario of sequential symmetry breaking \cite{FJM}, the scale 
$\Lambda_{F_{\rm ds}}$ is bounded  from the processes  
$K^+ \to \pi^+ \bar \nu \nu$. The latter competes  with $K^+ \to \pi^+\phi_{F_{\rm ds}}$ since the familon $\phi_{F_{\rm ds}}$ escapes detection due to its weak coupling to matter 
and ${\cal L}_{\rm eff} \simeq \frac{1}{\Lambda_{F_{\rm ds}}}\bar s_L \gamma_\mu d_L \partial^\mu \phi_{F_{\rm ds}}$, so that
the bound $\Lambda_{F_{\rm ds}} > 10^{8} \TeV$  is 
rather high \cite{feng}. The relation of this scale to the MFV-scale, which is roughly   
bounded $\Lambda_{\rm MFV} \geq 0.5\;{\rm TeV}$ from MFV-type operators \cite{MFV},
is a model dependent question.
If the breaking of flavour symmetry is decoupled from a lower physics scale, associated
with the stabilization of the Higgs mass for example, 
then the bounds do not directly apply. An example is a SUSY-GUT scenario, where 
it is assumed that flavour is generated or broken at some high scale and resides 
in so-called soft terms. Then operators of the MFV-type are generated when supersymmetry is broken and the MFV-scale is associated with the SUSY scale rather
than $\Lambda_{F_{\rm ds}}$.
Summarizing, 
the dynamics of spontaneous symmetry breaking suggests that in a (continuous) 
MFV-scenario
the generation of flavour are outside (current) experimental reach. As argued above this
does not exclude the observation of novel flavour effects due to an additional sector 
such as SUSY-GUT. 

In this paper we aim to ameliorate this situation by  replacing $G_F$  by a discrete symmetry.
Spontaneous breaking  of discrete symmetries do \emph{not} lead to Goldstone bosons e.g. \cite{SSBGF}. The absence of the latter in this framework  constitutes the \emph{primary} motivation of our 
work\footnote{Another alternative is to gauge the flavour symmetries, i.e. use
the Higgs-Englert-Brout mechanism. A dedicated analysis \cite{AFM} has been announced 
in reference \cite{FJM}. This was investigated in connection with family  symmetries 
some time ago \cite{higgs-mech}.}. 
The main remaining issue is then to investigate whether the reduced symmetries (or what discrete 
groups)  do provide enough protection  for a discrete TeV-scale  MFV-scenario.
Are the bounds on the coefficient $C_{\rm dMFV}$ in the TeV-scale range?

On the technical side this endeavour amounts to study the invariants of discrete subgroups. The reduced symmetry unavoidably leads  to further invariants
as compared to $G_F$. This bears as
a consequence that the flavour-mass basis transformation will become observable. 
The crucial question for discrete MFV is then at what level these new invariants are coming in. 
In this respect we find it useful to distinguish models where $\Delta F = 2$ operators arise 
from two $\Delta F = 1$ processes, as in the SM, and those where this is not the case.

Finally  we would like to stress to points. First, since we are following the effective field theory approach
there is no obvious connection to the scheme of constrained MFV \cite{CMFV}, which assumes
no new operators beyond those present in the SM.
Second,  there is no attempt made in this paper to explain
valuable textures of the Yukawa matrices, i.e. the question of what is flavour.
The symmetry is solely used to ensure that the Yukawa structure gouverns all flavour transition.
Our work is complementary to the field 
of flavour models with family symmetries, revived
by recent experimental information on neutrino masses and mixing angles (PMNS matrix).
Those models often involve discrete symmetries and  extended Higgs sectors  attempting to explain 
the origin of flavour hierarchies.
For a recent review and references on the subject, with emphasis on the neutrino 
sector, we refer the reader to the write-up \cite{review}.

The paper is organized as follows: 
In section \ref{sec:form} we state the problem in a more precise form.
Section \ref{sec:dg} summarizes some useful facts about groups and
gives an overview of the discrete SU(3) subgroups. 
In section \ref{sec:invariants} it is shown at what level new invariants necessarily arise
and which groups have the least invariants.
Section \ref{sec:physics} deals with the physical consequences of the previous findings
and proposes to distinguish flavour models according to generation mechanism of $\Delta F =2$
operators. In the epilogue we summarize our findings and reflect on the 
framework and its possible extensions.

\section{Formulation}
\label{sec:form}

\subsection{Minimal Flavour Violation}

In the SM  the quark masses and the CKM structure originate from the Yukawa Lagrangian, 
\begin{equation} 
\label{eq:yuk}
 {\cal L}_{\rm Y} = 
\bar Q_L H \yuk{D} D_R + 
\bar Q _L H_c \yuk{U} U_R \; + \; {\rm h.c.} \, ,
\end{equation}
 which breaks the flavour symmetry down to $U(1)_B$, c.f. Eq.~\eqref{eq:yuk_break}. 
The symbols $U_R$,  $D_R$  and $Q_L = (U_L,D_L)^T$ denote right and left handed SU(2)$_L$ singlets and
doublets respectively of up $U = (u,c,t)$ and down $D = (d,s,b)$ quarks.
The conjugate Higgs field is defined as $H_c = i \tau_2 H^*$.

It is observed that  the quark flavour symmetry,
\begin{equation}
\label{eq:Gq}
G_q = SU(3)_{Q_L} \times SU(3)_{U_R} \times SU(3)_{D_R} \;,
\end{equation}
can be formally restored by associating to the Yukawa matrices the following transformation properties:
\begin{equation}
\label{eq:trans}
\yuk{U} \sim (\R{3}, \Rb{3},\R{1})_{G_q}  \;, \qquad \yuk{D} \sim (\R{3},\R{1},\Rb{3})_{G_q} \; .
\end{equation}
In fact the flavour symmetry can even be further enhanced by two U(1) factors by appropriately assigning
U(1) charges to the quark fields and the Yukawa matrices. In our opinion 
there is some freedom in choosing them\footnote{
Any pair of U(1) charges for the fields 
$Q_L, U_R, D_R, \yuk{U}, \yuk{D}$ which leaves
\eqref{eq:yuk} invariant is in principle an option.
N.B. in reference  \cite{FJM} $U(1)_{U_R} \times  U(1)_{D_R}$ was chosen.}.

An effective theory constructed from the SM fields and the Yukawa matrices
is then said to obey  the principle  of \emph{Minimal Flavour Violation} \cite{MFV}, 
if the higher dimensional operators are invariant under $G_q$ and CP\footnote{
The latter condition might be relaxed by allowing for arbitrary CP-odd phases in the coefficients $C_{\rm MFV}$ of the effective operators.
This has been done for example for the  MSSM in reference \cite{kaons}. One could go even further and assume strong phases as well, which could be
due to low energy degrees of freedom, such as the ones studied in the unparticle scenario \cite{Georgi1}. Working in the  MFV scenario we though implicitly assume that the new structure does not involve new light degrees of freedom.}. Operators with $\Delta F = 1,2$ then assume the following form \cite{MFV}
\begin{alignat}{1}
\label{eq:DeltaF}
O^{\Delta F = 1'} &=  (\bar D_L \yuk{U} \yuk{U}^\dagger \yuk{D} \, \sigma \!\cdot\! F  D_R)  \;  ,
\nonumber \\
O^{\Delta F = 1} &=  (\bar D_L \yuk{U} \yuk{U}^\dagger \gamma_\mu D_L)  \, \bar D_L \gamma^\mu D_L  \;,\nonumber \\
O^{\Delta F = 2} &=  (\bar D_L \yuk{U} \yuk{U}^\dagger \gamma_\mu D_L)^2
\;,
\end{alignat}
in the left handed $D$-sector.  The symbol $\sigma \!\cdot\! F
= \sigma_{\mu\nu} F^{\mu\nu}$ denotes  the contracted electromagnetic tensor, $2 D_{L/R} \equiv (1 \pm \gamma_5) D$ and we would like to add 
that notation $\Delta F =1'$, used here, is rather non-standard. In the remainder of this paper we shall omit the explicit indication of the $\gamma$-matrices.
Transitions to right handed quarks demand substitutions of the 
form $D_L \to \yuk{D} D_R$ etc, by virtue of $G_q$-invariance  (\ref{eq:Gq},\ref{eq:trans}), leading
to the well-known phenomenon of chiral suppression.
The operators in the $U$-sector are simply obtained by interchanging the role of the $U$ and $D$ families.
Let us parenthetically note that  the predictivity of MFV \cite{MFV} 
in the $D$ sector is 
in large parts due to the 
the fact that the top is much heavier than the other $U$-quarks
\begin{equation}
\label{eq:Dtrans}
\left( \yuk{U} \yuk{U}^\dagger \right)_{ij} = \left( V^\dagger {\rm diag}( y_u,y_c,y_t )^2  V \right)_{ij} 
\simeq y_t^2  V^*_{3i} V_{3j} \; ,
\end{equation}
where $V = {\cal U}_L^\dagger {\cal D}_L$ is the CKM matrix, resulting from the bi-unitary diagonalization of the 
\begin{equation}
\label{eq:biunitary}
U_{L} \to {\cal U}_{L} U_{L}\;, \quad U_{R} \to {\cal U}_{R} U_{R}\;, \quad  D_{L} \to {\cal D}_{L} D_{L} \;, \quad  D_{R} \to {\cal D}_{R} D_{R}  \; ,
\end{equation}
Yukawa matrices.
The masses are related to the Yukawa couplings as $\sqrt{2} m_i = v y_i$
(with $v \simeq 246 \GeV$) and it is worth noting that in the limit of
degenerate masses  the GIM mechanism reveals itself through:
$(Y_U^\dagger Y_U)_{ij} \to \delta_{ij} $.

\subsection{Discrete Minimal Flavour Violation}

Replacing  the continuous flavour symmetry with a discrete flavour symmetry requires the following additional information
or assumptions:
\begin{alignat}{2}
\label{eq:discrete}
&a)    \quad  \text{The  group} \quad & & 
D_q  = {\cal D}_{Q_L} \times {\cal D}_{U_R} \times {\cal D}_{D_R} \subset G_q \;, \quad {\cal D} \subset SU(3) \;, \nonumber  \\
&b)  \quad  \text{The representation}  \quad & &  R_3({\cal D}_{Q_L}) \text{ (3D irrep of families )}  \;,  \nonumber  \\
&c)  \quad  \text{(possibly) Yukawa expansion}  \quad & &  \yuk{U/D} 
\to  \kappa \yuk{U/D} \;, \quad \kappa \in \mathbf{R} \;,
\end{alignat}
Since the three families have to transform in a 3D irreducible representation (irrep) 
this leads us naturally to study the discrete SU(3) subgroups, which we denote by 
the symbol ${\cal D}$. The irrep has to be specified since some groups have more than one of them.   The
reduced symmetry a)  leads to new invariants and therefore renders the transition matrices \eqref{eq:biunitary} observable. We will argue in subsection \ref{sec:mass}
that this gives rise to a rather anarchic pattern of flavour transitions.
This can be controlled by assuming a hierarchy
in the Yukawa expansion\footnote{ 
Such a notion has for example been introduced in 
reference \cite{genMFV}. The authors use
the notation $\epsilon_{u,d} Y_{u,d}$ and distinguish 
the case $\epsilon_{u,d} \ll 1$ linear MFV and 
$\epsilon_{u,d} \sim {\cal O}(1) $ non-linear MFV. In the latter case a non-linear $\sigma$-model techniques are 
imperative \cite{FMtop,genMFV}.}. 
In a perturbative-type model for example 
the operators with
several Yukawa  insertion might originate from 
higher dimensional operators suppressed by some 
high scale $\Lambda$ and could mean
$\kappa \simeq v/\Lambda$ if the Yukawa assume
a vacuum expectation value (VEV) around the electroweak scale.
A rough but conservative estimate in subsection 
\ref{sec:superweak} indicates that for $\kappa 
\simeq 0.2$
$C_{\rm dMFV}$ has the same bounds as $C_{\rm MFV}$.
In this paper we shall not discuss the U(1) factors,
e.g. \eqref{eq:yuk_break}, any further. We can think
of them as being replaced by a discrete 
$\mathbb{Z}_n$ symmetry, $D_q \to D_q \times \mathbb{Z}_n  ..$
and they do not play a role in the type of invariants
we are considering\footnote{ 
In principle though, one could think of embeddings
where they play a more subtle role, c.f. epilogue.}.
Generally the embedding could play a more subtle role.
First there is freedom of embedding  ${\cal D}$ into SU(3). 
We will discuss
this issue in section \ref{sec:mass} where it is argued that the obersvability of the rotation matrices \eqref{eq:biunitary} can be
suppressed, modulo $V_{\rm CKM}$, for certain groups by a suitable embedding. Second, we would like to draw the reader's attention to the assumption of $D_q$ being embedded as a direct product of the discrete SU(3) subgroups a) into $G_q$. We shall comment
on it in the epilogue.

\section{On (discrete) groups}
\label{sec:dg}

In this section we shall first state a few useful facts about invariants, groups and representations, which is at the heart of
this paper. Then we shall say a few 
things about the classification of discrete SU(3) subgroups.

\subsection{Useful facts}
\label{sec:useful}

Consider irreps $\{ \R{A}, \R{B}, \R{C}, ... \}$
of some group, continuous or discrete, and denote the explicit vectors of the irreps by 
${\cal V} = \{ a_r, b_s, c_t, ... \}$.
By the orthogonality theorem the number of times that the identity appears in the Kronecker product, denoted by $n_1$,
\begin{equation}
\R{A} \times \R{B} \times  \R{C} \times \dots = n_1 \, \R{1} + \dots  \;,
\end{equation}
is \emph{equal} to the number of independent invariants
\begin{equation}
\I_n = \I_n^{rst..}\, a_r  b_s  c_t .. \;,   \qquad n = 1 \dots n_1 \;,
\end{equation}
that can be built out of the set ${\cal V}$ specified above. Throughout this paper repeated indices are thought to be summed over.
It is this statement that we shall use shortly 
for our main result. Before going on we would like to mention another fact, peculiar to discrete groups: 
The number of elements of the group is equal,
\begin{equation}
\label{eq:fund}
|{\cal D}| = \sum_{i \in \text{irreps}} {\rm dim} (R_i({\cal D}))^2 \;,
\end{equation}
to the sum of the squared dimensions of all irreps.

\subsection{Discrete SU(3) subgroups}
\label{sec:DSU(3)}

The discrete subgroups of SU(3) were classified a long time ago \cite{Milleretal}
and further analyzed as alternatives to SU(3)$_F$ in the context of the eightfold way  
\cite{FFK}. Explicit representations and Clebsch-Gordan coefficients were systematically worked out
in a  series of papers around 1980 \cite{BW,BLW1}, partly motivated by the usage of discrete SU(3) subgroups
as an approximation to $SU(3)_{\rm colour}$ in lattice QCD; and further elaborated very recently 
\cite{simple_discrete,Delta3n2,Delta6n2} in the context  of 
family symmetries. 
That the catalogue of \cite{FFK} as compared
to \cite{Milleretal} is not complete already surfaced
in the 1980 \cite{BLW1,response} and
has recently been discussed more systematically in a  remarkable diploma thesis \cite{Ludl}.

The discrete subgroups of SU(3) are of two kinds. 
The first type consist of the analogue of crystal groups of which we list here the maximal subgroups: 
$\Sigma(168)$,  $\Sigma(360 \varphi)$ and $\Sigma(216 \varphi)$.
The factor $\varphi$ can in general 
either be one or three depending on whether the 
center of SU(3) is divided out or not. For the maximal subgroups it is three\footnote{The further groups that 
are listed in \cite{FFK} are all subgroups:
$\Sigma(60) \subset \Sigma(360\varphi)$, 
$\Sigma(36 \varphi) \subset \Sigma(72 \varphi) \subset \Sigma(216 \varphi)$ for $ \varphi = 1,3$ and of course 
generally $\Sigma(n) \subset \Sigma(n \varphi)$ for 
$n = 36,72,216,360$ \cite{FFK,Ludl}. 
Once we have established an interesting candidates we shall proceed to have a look at its maximal subgroups.}.
The second kind are the infinite sequence of groups, sometimes called ``dihedral like'' or ``trihedral'', 
$\Delta(3n^2)   \iso (\z{n} \times \z{n})  \rtimes \z{3}$ and 
$\Delta(6n^2)  \iso (\z{n} \times \z{n})  \rtimes S_3 $ for $n \in {\mathbb Z}$. The symbol ``$\iso$'' stands for isomorphic and 
``$\rtimes$'' denotes a semidirect product. 
The largest irreps of $\Delta(3n^2)$ and 
$\Delta(6n^2)$ are 3-, respectively 6-dimensional; independent of $n$. 
Note, for all groups the number denotes the order of the group, i.e. the number of elements. 
In appendix \ref{app:Dgroups} we argue that the $(D)$-groups recently emphasized in \cite{Ludl}, or more precisely the six-parameter
$D(n,a,b;d,r,s)$ matrix groups,  
are  subgroups of $\Delta(6g^2)$, where
$g$ is the  lowest common multiple of $n$, $d$ and $2$. We therefore do 
not need to discuss them separately.
Further aspects of some groups and some of their invariants 
are discussed in the appendix \ref{app:examples}.

We would like to add that since this paper has been first published further progress
has been achieved by P.Ludl \cite{Ludl2011}, 
who showed that topologically the $(C)$ and $(D)$-groups 
correspond to $(C)  \iso (\z{n} \times \z{n'})  \rtimes \z{3}$ and 
$(D)  \iso ((\z{n} \times \z{n'})  \rtimes \z{3}) \rtimes  \z{2}$.

\section{Invariants}
\label{sec:invariants}

The study of invariants $\I$ is the main ingredient of this paper since the effective Lagrangian approach. 
In the absence of the knowledge of the dynamics of the underlying model the effective Lagrangian 
assumes the following form
\begin{equation}
\label{eq:Leff}
{\cal L}^{\rm eff} = \sum_n C^n_{\rm dMFV}   \left(\I_n(\text{Quarks},\text{Yukawas} \right) + h.c)\, , \qquad  C^n_{\rm dMFV}=  \frac{c_n}{  \Lambda^{\text{dim}(\I_n) -4} } \;,
\end{equation}
where the dimension of the operator (invariant) brings in a certain hierarchy in the infinite sum above.  As previously mentioned the association of 
$C^n$ with the scale of new structure is generally obscured by 
loop factors, mixing angles and renormalization group effects.
Finding the invariants is equivalent to finding the constant tensors
of the symmetry group. All objects are either in a $\R{3}$ or the corresponding $\Rb{3}$ representation for which we
write lower and upper indices, e.g.\footnote{In this notation the basic constant tensors of SU(3) are $\I^{(1,1)} \sim  \delta_a^{\phantom{x}b}$
, $\I^{(0,3)} \sim \epsilon^{abc}$
and $\I^{(3,0)} \sim \epsilon_{abc}$ c.f.\cite{groups}.}  
\begin{equation}
\I^{(m,n)} \; \sim \; \I_{\,a_1 .. a_m}^{\,b_1 .. b_n}  \;,
\end{equation}
as is common practice in the literature e.g. \cite{groups}. Non-constant tensors will be denoted by  ${\cal T}^{(m,n)}$.
In principle this tensor classification is not sufficient for our general problem since there are three different group factors \eqref{eq:discrete}. 
It will though prove sufficient here to contract the other indices\footnote{
A refined treatment is only necessary when there
are new $\I^{(2,2)}$ invariants and those groups are not of interest to us anyway as explained in the text.}.
We therefore contract the  ${\cal D}_{U_R}$ index and directly write
 \begin{equation}
 \label{eq:tran}
 (\tran_U) _a^{\phantom{x}r} \equiv ( \yuk{U} \yuk{U}^\dagger)_a^{\phantom{x}r}\;, \qquad     ( \tran 
 \in {\cal T}^{(1,1)}) \;.
\end{equation}
We shall often drop the subscript $U$ when there is no reason for confusion.
In the reminder we shall use the following notation for left handed down quarks
\begin{equation}
\label{eq:para3}
D_L = (d_L , s_L, b_L) \to D_i = (D_1,D_2,D_3) \;, \qquad ( D_i \in {\cal T}^{(1,0)}) \;.
\end{equation}
The operators in Eq.~\eqref{eq:DeltaF} are associated with invariants of the form:
\begin{alignat}{2}
& \I^{(2,2)}_n = (\I_n)^{ab}_{rs}  \, 
\left( \bar D^r  \tran_a^{\phantom{x}s} D_b \right)  \qquad & & \Delta F = 1' \;,
\nonumber \\ 
& \I^{(3,3)}_n =  (\I_n)^{abc}_{rst}  \, 
\left( \bar D^r  \tran_a^{\phantom{x}s} D_b \right) \bar D^t D_c
  \qquad & & \Delta F = 1 \;, \nonumber \\ 
\label{eq:generic2}
&  \I^{(4,4)}_n =  (\I_n)^{abcd}_{rstu}  \, 
\left( \bar D^r  \tran_a^{\phantom{x}s} D_b \right) \, \left( \bar D^t  \tran_c^{\phantom{x}u} D_d \right) \qquad & & \Delta F = 2  \;.
\end{alignat}
Note that there are groups where $\Delta F =2$ operators are possible with $\I^{(2,2)}$ 
but those structures are definitely too far away from MFV to be of any interest to us. 
The group $\Sigma(60) \iso A_5$ is an example which is discussed in appendix \ref{app:crystal}.

\subsection{No $\R{27}$ ! -- On the necessity of new invariants at the $\I^{(4,4)}$ level}
\label{sec:necessity}

In this subsection we will present a general argument that there are necessarily further invariants for $\I^{(4,4)}$, corresponding to
the generic $\Delta F = 2$ transition \eqref{eq:generic2},
as compared to  the SU(3) case.
Note the $\Delta F = 2$ operator \eqref{eq:DeltaF} 
of $G_q$ is obtained for
$(\I_{1})^{abcd}_{rstu}  \equiv \delta_r^{\phantom{x}a}
\delta_s^{\phantom{x}b} \delta_t^{\phantom{x}c}
\delta_u^{\phantom{x}d}$.
The problem in Eq.~\eqref{eq:generic2} reduces to finding the invariants
of the following Kronecker product
\begin{equation}
\label{eq:basic}
K^{\cal D_{Q_L}} = \left( \Rb{3} \times  \R{3}  \times  \Rb{3} \times \R{3} \right)_s \times \left( \Rb{3} \times
  \R{3}  \times  \Rb{3} \times \R{3} \right)_s  = n_1 \R{1}  + ...  \; ,
\end{equation}
where $s$ stands for the symmetric part and can be justified as follows:
Since the Kronecker product is associative we may choose an ordering adapted to the symmetries in Eq.~\eqref{eq:generic2}. The most economic way
is to tensor the quarks and the Yukawas separately
for which only the symmetric part, indicated above,
is needed.
Let us first look at the invariants that can be generated in the case where
${\cal D}_{Q_L} \to SU(3)$:
\begin{equation}
K^{SU(3)} = \left( (\R{1} + \R{8}) \times  (\R{1} + \R{8})\right)_s \times  \left((\R{1} + \R{8}) 
\times  (\R{1} + \R{8}) \right)_s
\end{equation}
We shall focus on the product of the four $\R{8}$'s,
\begin{equation}
\label{eq:symmetric}
K^{SU(3)} = \left(\R{8} \times \R{8}\right)_s \times 
\left(\R{8} \times \R{8}\right)_s + ...  \sim \left( \bar D T^A D \times \bar D T^B D \right)_s  \times 
\left( {\rm tr}[ \tran T^C]   \times{\rm tr}[ \tran T^D]   \right)_s + ... \;.
\end{equation}
The Clebsch-Gordan coefficients of the $\R{8}$ are, of course, just the generators
$T^A$, $A= 1..8$ of the SU(3) Lie Algebra. The Kronecker product of the 8D irrep in SU(3) reads, e.g. \cite{Slansky},
\begin{equation}
\label{eq:27}
(\R{8} \times  \R{8})_{SU(3)} = 
(\R{1} + \R{8} + \R{27})_s  + (\R{8} + \R{10} + \Rbb{10})_a  \;.
\end{equation}
As argued above only  the symmetric part is needed.
Eq.~\eqref{eq:27} is telling us on the one hand that the decomposition 
in \eqref{eq:symmetric} will lead to three different invariants but more importantly
it tells us how the discrete subgroup ${\cal D}_{Q_L}$ has to decompose in order not to generate further invariants.
A necessary condition for an identical decomposition is that the discrete group contains 
a $\R{27}$.
The trihedral groups  $\Delta(3n^2)$ and $\Delta(6n^2)$ are not in this category since their largest
irreps are at most 3-, respectively 6-dimensional. The same applies
to the $(C)$-groups \cite{Ludl} 
and $(D)$-groups, c.f. appendix \ref{app:Dgroups}, since they can be embedded into
$\Delta(3g^2)$ and $\Delta(6g^2)$ for an appropriate $g$. Thus we are left with the groups 
of the crystallographic type.
Going through the character tables  in \cite{FFK} and the more recent work \cite{Ludl} we realize that there is \emph{no} discrete subgroup 
of SU(3) which has a 27D irrep! 
Note, on even more general grounds  that ${\rm dim}(\R{27})^2 = 
729$ almost saturates  the relation in Eq.~\eqref{eq:fund} and leaves  $|\Sigma(360 \varphi)|= 1080$ as the only hypothetical 
candidate among the crystal-like groups.

\subsection{Groups with no new invariants at the $\I^{(2,2)}$ ($\I^{(3,3)}$) level}
\label{sec:nonew}
In this subsection we investigate the structure of the invariants $\I^{(2,2)}$, 
which correspond to  $\Delta F = 1'$ transitions \eqref{eq:generic2}. 
With the same reasoning as above this reduces to analyze,
\begin{equation}
\label{eq:kron3333}
K^{\cal D_{Q_L}} = \left( \Rb{3} \times  \R{3}  \times  \Rb{3} \times \R{3} \right) = n_1[{\cal D}_{Q_L}] \,
\R{1} + ... \;,
\end{equation}
the number 
of invariants that can be formed from the Kronecker product \eqref{eq:kron3333}.
According to subsection \ref{sec:useful}  the number of invariants 
in SU(3) equals the number of invariants of
the discrete subgroup, i.e.  $n_1[{\cal D}_{Q_L}] 
= n_1[SU(3)] $ if we can choose a $R_3( {\cal D}_{Q_L})$ such that
\begin{equation}
\label{eq:8}
\left( \R{3} \times \Rb{3}\right)_{{\cal D}_{Q_L}} = 
\R{1} + \R{8} \; .
\end{equation}
Once more the trihedral groups  $\Delta(3n^2)$, $\Delta(6n^2)$ can be excluded immediately since
they posses maximally 3-, respectively 6-dimensional irreps. 
It turns out that all  three maximal crystal-like subgroups 
 $\Sigma(168)$,    $\Sigma(216 \varphi)$ and $\Sigma(360 \varphi)$ (recall $\varphi=3$) are in the category of \eqref{eq:8}
and do therefore not generate any further $\I^{(2,2)}$ invariants.
Below we shall briefly discuss the representations and some other relevant aspects of the candidates.
\begin{itemize}
\item[A)] $\Sigma(168) \iso {\rm PSL}(2,\mathbb{F}_7) \iso 
{\rm GL}(3,\mathbb{F}_2)$ \cite{simple_discrete} is discussed in some more detail in appendix \ref{app:crystal} as well.
In what follows we shall permit ourselves to present the irreps 
in a compact way, though not unambiguous in a strict sense, via the relation \eqref{eq:fund} \,
\begin{equation}
\label{eq:168}
|\Sigma(168)| = 168 = |\R{1}|^2 + 2   |\R{3}|^2 + 
|\R{6}|^2 + |\R{7}|^2 + |\R{8}|^2 \,,
\end{equation}
Explicit representation matrices 
 can be found in \cite{simple_discrete} and the one 3D irrep  satisfies \eqref{eq:8}.
\item[B)] $\Sigma(216 \varphi)$:  Eq.~\eqref{eq:fund} reads \cite{Ludl}
\begin{equation}
|\Sigma(216 \varphi)| = 3\cdot 216 = 3 |\R{1}|^2 + 3 |\R{2}|^2 + 7 |\R{3}|^2
+ 6  |\R{6}|^2 + 3 |\R{8}|^2 + 2 |\R{9}|^2   \;.
\end{equation}
Out of the seven 3D irreps one is not faithful\footnote{It is isomorphic to $A_4$ \cite{Ludl}, which is a  popular group in attempts to explain tri-bi maximal mixing, e.g.\cite{review}.} 
and out of the three eight dimensional irreps two are complex and therefore 
not of interest. The other six 3D irreps come  in complex conjugate pairs. In the notation of \cite{Ludl}
\begin{equation}
\label{eq:216phi}
\Sigma(216 \varphi): \quad  \R{3}_2 \times \R{3}_6 = \R{3}_3 \times \R{3}_5 
= \R{3}_4 \times \R{3}_7  = \R{1}_1 + \R{8}_1 \;,
\end{equation}
are  the pairings \`a la Eq.~\eqref{eq:8}.
\item[C)] $\Sigma(360 \varphi)$: Eq.~\eqref{eq:fund} reads \cite{Ludl}
\begin{equation}
|\Sigma(360\varphi)| = 3\cdot 360 =   |\R{1}|^2 + 4 |\R{3}|^2 + 2 |\R{5}|^2
+ 2 |\R{6}|^2 + 3|\R{8}|^2 + 3|\R{9}|^2  
+ |\R{10}|^2 + 2 |\R{15}|^2  \;.
\end{equation}
In the notation of \cite{Ludl} the interesting Kronecker products are
\begin{equation}
\label{eq:360phi}
\Sigma(360 \varphi): \quad  
\R{3}_1 \times \R{3}_4  = \R{1}_1 + \R{8}_1 \;, 
\qquad \R{3}_2 \times \R{3}_3 = \R{1}_1 + \R{8}_2 \;,
\end{equation}
which implies that both 8D irreps are real and both
are valid candidates for our problem.
\end{itemize}
Of course the question of whether any subgroups 
of $\Sigma(168)$,  $\Sigma(360 \varphi)$,  $\Sigma(216 \varphi)$ are in the category \eqref{eq:8} is a relevant question here.
Some, of the smaller groups, can be excluded on grounds
of their order since by virtue of \eqref{eq:fund}, Eq.~\eqref{eq:8}
demands
\begin{equation}
\label{eq:74}
\text{Order} \geq 74 =  8^2 + 3^2 + 1^2  \;.
\end{equation}
The group $\Sigma(168)$ has the permutation group $S_4$ and
the Frobenius group   $ \z{7}   \rtimes \z{3}$ as maximal subgroups of which both can be discarded
since their  order, $|{\cal S}_4| = 24$ \& 
$ |\z{7}   \rtimes \z{3}|=21$, does not satisfy \eqref{eq:74}. In the case of $\Sigma(216\varphi)$ we are aware of two maximal subgroups, 
$\Sigma(216) = \Sigma(216\varphi)/ \z{3}$ and $\Sigma(72 \varphi)$. The first one has a single 3D irrep which decomposes as $\R{3} \times \R{3} = \R{1} +
\R{1}' + \Rb{1}' + 2 \cdot \R{3}$ and is therefore not suitable. The second one: 
\begin{itemize}
\item[D)] $\Sigma(72 \varphi)$:  Eq.~\eqref{eq:fund} reads \cite{Ludl}
\begin{equation}
|\Sigma(72 \varphi)| = 3\cdot 72 =   4 |\R{1}|^2 +  |\R{2}|^2 + 8 |\R{3}|^2 + 2 |\R{5}|^2 + 2 |\R{6}|^2 + |\R{8}|^2  \;,
\end{equation}
and the eight 3D irreps fall into four complex conjugate pairs as follows \cite{Ludl}
\begin{equation}
\label{eq:72phi}
\Sigma(72 \varphi): \quad 
\R{3}_1 \times \R{3}_2 = \R{3}_3 \times \R{3}_4 
= \R{3}_5 \times \R{3}_6  =  \R{3}_7 \times \R{3}_8 =
\R{1}_1 + \R{8}_1 \;,
\end{equation}
and we therefore have four choices for a pair of 
irreps. 
\end{itemize}
In the case of $\Sigma(360 \varphi)$ we are
aware of the two maximal subgroups 
$\Sigma(60) \iso A_5$ and 
$\Sigma(360) \iso \Sigma(360\varphi)/ \z{3}$ of which the former can be excluded by virtue of \eqref{eq:74}
and the latter does not admit a  
3D irrep\footnote{The same thing happens in the continuum; the 
$SU(2)/\mathbb{Z}_2 \iso SO(3)$ does not admit 2D representations.} 
as can be inferred from the character table e.g. \cite{FFK} and is therefore not suitable.  

In order to count the number of invariants up to $\I^{(4,4)}$ it is sufficient to know  the $\R{8} \times \R{8}$ 
Kronecker produtcs.  
We shall list those for  SU(3) \eqref{eq:27}, $\Sigma(168)$  \cite{simple_discrete} 
$\Sigma(72 \varphi)$,  $\Sigma(216 \varphi)$ and $\Sigma(360 \varphi)$.
The latter three have been computed from the
character tables given in reference \cite{Ludl}:
\begin{alignat}{2}
\label{eq:88}
& (\R{8} \times  \R{8})_{SU(3)} \; &=& \;
\left(\R{1} + \R{8} + \R{27} \right)_s  + \left(\R{8} + \R{10} + \Rbb{10}\right)_a \;, \nonumber \\
&  (\R{8} \times  \R{8})_{\Sigma(168)} \; & = & \;
\left(\R{1} + \R{8} + ( 2 \cdot \R{6}  + \R{7} + \R{8}) \right)_s  + \left(\R{8} + (\R{3} +\R{7}) + (\Rb{3} + \R{7})\right)_a \;, \nonumber \\
&  (\R{8} \times  \R{8})_{\Sigma(72\varphi)} \; & = & \;
\left(\R{1}_1 + \R{8} + (\R{1}_2 + \R{1}_3 + \R{1}_4  + 3 \cdot \R{8}) \right) _s  + \left(\R{8} +  (\R{8} + \R{2}) + (\R{8} + \R{2})  \right)_a \;, \nonumber \\
&  (\R{8} \times  \R{8})_{\Sigma(216\varphi)} \; & = & \;
\left(\R{1} + \R{8}_1 +  (\R{3}_1 + \R{8}_1  + \R{8}_2 + \R{8}_3  ) \right) _s  + \left(\R{8}_1 +  (\R{8}_2 + \R{2}_2) + (\R{8}_3 + \R{2}_3)  \right)_a \;, \nonumber \\
&  (\R{8}_1 \times  \R{8}_1)_{\Sigma(360\varphi)} \; & = & \;
\left(\R{1}_1 + \R{8}_1 + (\R{5}_1 + \R{5}_2 + \R{8}_2  +  \R{9}_1 \right) _s  + \left(\R{8}_1 +  \R{10}  + \R{10} )  \right)_a \;.
\end{alignat}
The brackets indicate the branching rules. The a priori unclear pairings 
$(\R{8}_2 + \R{2}_2)_{\Sigma(216\varphi)}$  and $(\R{3} +\R{7})_{\Sigma(168)}$ can be inferred from reference \cite{branch}.
The product 
$ (\R{8}_2 \times  \R{8}_2)_{\Sigma(360\varphi)}$ is obtained from the one in \eqref{eq:88} by interchanging the subscripts $1 \leftrightarrow 2$.
\begin{table}[h]
\begin{center}
\begin{tabular}{l | r |  c | c | c c | c c c |}
group &  order & pairs $(\R{3},\Rb{3})$ & $\I^{(2,2)}$ & $\I^{(3,3)}$ & $\I^{(3,3)}_{2,1}$  & $\I^{(4,4)}$ & $\I^{(4,4)}_{3,1}$ & $\I^{(4,4)}_{2,2}$  \\[0.1cm]
\hline
& & & & & & & & \\[-0.3cm]
SU(3)                                     & $\infty$ \quad   & 1& 2  & 6   & 5 & 23 & 15 & 14 \\
$\Sigma(360\varphi)$            & 1080  \quad  & 2 & 2  & 6  &  5 &  28 & 18 & 17 \\
$\Sigma(216\varphi)$            & 648  \quad &3 &2  & 7  &  6 &  40 & 27 & 23 \\
$\Sigma(168)$                       & 168  \quad &1 & 2  & 7   & 6 & 44 & 29 & 25 \\
$\Sigma(72\varphi)$              & 216  \quad &4 & 2  & 11 & 8 &  92 & 55 & 43 \\
\end{tabular}
\end{center}
\caption{\sl \small Number of complex conjugate pairs and number of invariants for tensors of type $\I^{(n,n)}$, whose definition 
can be inferred from Eq.~\eqref{eq:generic2}. 
The subscripts $x,y$ indicate symmetrizations of $x$ and $y$ pairs of 
$\R{3},\Rb{3}$ indices. $\I^{(3,3)}_{2,1}$ and $\I^{(4,4)}_{2,2}$ correspond
to the (symmetric) contractions of
$\Delta F = 1$ and $\Delta F =2$ in \eqref{eq:generic2}.} 
\label{tab:candidates}
\end{table}
To this end we shall give an overview of the number of invariants for $\I^{(2,2)}$, $\I^{(3,3)}$ and $\I^{(4,4)}$ in Tab.~\ref{tab:candidates}.
The number of 3D complex conjugate pairs are also listed. The symmetrized tensors $\I_{x,y}$ are explained in the caption.


\section{Back to physics}
\label{sec:physics}
\subsection{Flavour to  mass basis -- New invariants lead to flavour anarchy}
\label{sec:mass}

By switching from the
flavour basis to the physical (mass) basis we employ bi-unitary transformations \eqref{eq:biunitary} 
in a
$U(3)_{U_L} \times U(3)_{D_L}  \times U(3)_{U_R} \times 
U(3)_{D_R}$ space.
In the SM this group is broken down to $U(3)^3$, c.f.
Eq.~\eqref{eq:Gq}\footnote{In this section we will take a cavalier attitude towards the question of the proper U(1) factors.},
by the  SU(2)$_L$ gauge group, rendering  
$V_{\rm CKM} = {\cal U}_L^\dagger {\cal D}_L$ observable.
By choosing a discrete symmetry $D_q$ \eqref{eq:discrete}
the group is  further  broken down and  this will 
in \emph{general} render the transformation matrices \eqref{eq:biunitary} observable\footnote{From another viewpoint it is the absence of the Goldstone bosons, in the approach followed here, that leads to further observable parameters. The latter are 
in one to one correspondence with  the reduction of observable
parameters. Counting in quark sector:  
$18\cdot 2$ Yukawa parameters minus 26 Goldstones gives
4 CKM parameters and 6 quark masses \cite{FJM}.
} at the order in the Yukawa expansion where the invariants of 
the groups ${\cal D}$ and SU(3) differ.

One can take different points of view here. 
From a certain perspective the misalignement of the flavour and mass basis is simply observable and the result
has to be accepted. We can though push the bar and take a more active viewpoint and ask the question:
Given arbitrary Yukawa matrices is there an embedding, 
of the  discrete group ${\cal D}$ into SU(3) which allows us to choose ${\cal U}_R \simeq 1$, ${\cal D}_R \simeq 1$ and  either ${\cal U}_L \simeq 1$ or ${\cal D}_L \simeq 1$? Note that ${\cal U}_L $ and ${\cal D}_L$ can be interchanged via 
the CKM matrix. This question is investigated 
in appendix \ref{app:embedding} and the answer 
is no for the crystal  groups but  
appears to be affirmative for the trihedral  
group $\Delta(3n^2)$ and $\Delta(6n^2)$ for adjusted $n$ and choice of 3D irrep.
Although this in principle allows to make the new angles and
phases arbitrarily small it is a fact that non-SU(3) invariants 
by themselves lead to new flavour patterns.
 Let assume that ${\cal U}_R \simeq 1$, ${\cal D}_R \simeq 1$ and  ${\cal U}_L \simeq 1$, implying
${\cal D}_L \simeq V_{(CKM)}$, then the $\Delta(3n^2)/\Delta(6n^2)$ invariant (\ref{eq:3n2_I2},\ref{eq:6n2_I})
becomes
\begin{eqnarray}
\I^{(2,2)}_2  &\sim&  (\bar D^1 D_1  \bar D^2 D_2 + \bar D^2 D_2  \bar D^3 D_3 + 
\bar D^1 D_1  \bar D^3 D_3) \;, \nonumber \\
\label{eq:hm}
 &\to& \left( \bar d \,  (V_{ud}^*  V_{us})  \, s_L \right) \left(  \bar d\,  (V_{cd}^* V_{cs}) \, s_L \right) + .... \quad  \leftrightarrow \quad
{\cal L}_{\rm eff}^{\Delta S= 2} \sim {\cal O}(\lambda^2) \;,
\end{eqnarray}
becomes  a  $\Delta S= 2$ operator, at second order in the Wolfenstein parameter ($\lambda \simeq \cos(\theta_C) \simeq 0.22$), in the mass basis.
 Whereas in the MFV scenario
this transition is gouverned by $|V^*_{ts} V_{td}|^2 \sim {\cal O}( \lambda^{10})$.
This seems rather anarchic.
New invariants therefore spoil predictivity and it seems desirable to track or control them in some way.

\subsection{If $(\Delta F= 2) \iso (\Delta F =1) \times (\Delta F =1)$ then $\I^{(4,4)} \to \I^{(2,2)}\I^{(2,2)}$ }
\label{sec:superweak}

The view that 
in  a quantum field theory any term, not forbidden by symmetry,  emerges dynamically 
is deeply rooted, e.g. \cite{weinberg1}. 
The crucial pragmatic question for our approach
is what are the bounds on the coefficients $C_{\rm dMFV}$. 
We will find it useful to divide models in certain classes and reflect on a few specific examples.

\begin{figure}[h]
 \centerline{\includegraphics[width=5.2in]{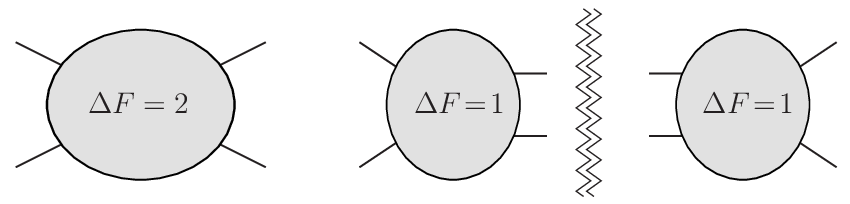}
 }
 \caption{\small (Left) ``family irreducible'': generic $\Delta F =2$ process
 (Right)  ``family reducible'': $\Delta F =2$ process can be disconnected into two  $\Delta F =1$ parts. There is no horizontal or family charge flowing from the left to the right. The ``family reducible'' property is a \emph{sufficient} property for a TeV-scale discrete MFV scenario for any of 
the groups Eq.~\eqref{eq:thefamousfour}.}
\label{fig:DeltaF12}
\end{figure}

In the case where there is no suppression of higher
order terms, other than the Yukawa expansion
itself, the results of subsection \ref{sec:necessity}  indicate that at the $\Delta F = 2$ level \eqref{eq:generic2} new invariants of the 
type $\I^{(4,4)}$  (could) emerge. This seems rather 
dangerous at first sight because the results of the 
previous subsection imply that new invariants upset
the flavour structure \emph{and} predicivity  since the mass-flavour basis transformation becomes observable.
As hinted at above it would appear too hasty to conclude that no discrete flavour group is suitable. 
The generation mechanism of  $\Delta F = 2$ 
operators has to be reflected upon.
In rather general terms we may want to distinguish the two cases where the $\Delta F =2$ process 
is generated via two
subsequent $\Delta F = 1$ parts and where
this is not the case\footnote{One is tempted to say somewhat in the spirit 
of the phenomenological superweak model for CP-violation \cite{superweak}, bearing in mind though that not all features such as for example 
the reality of the CKM matrix are relevant here.}. 
We  shall call the former  ``family irreducible'' 
and the latter ``family reducible'',  c.f. Fig.~\ref{fig:DeltaF12}.  
The SM or the R-parity conserving 
Minimal Supersymmetric Standard Model (MSSM) c.f. Fig.~\ref{fig:examples_reducible}, as presumably many perturbative models, are of the ``family reducible'' type. 
The composite technicolor model of reference 
\cite{georgi_chivukula}
cannot be claimed, in the absence of the understanding of the non-perturbative dynamics of
the preon-confinement, to be in the 
``family reducible'' class.

\vspace{0.2cm}
The ``family irreducible'' property (c.f. Fig.~\ref{fig:DeltaF12})  is a \emph{sufficient} condition for a TeV-scale discrete MFV scenario if the global flavour group $D_q$ is built from the following crystal-like  groups
\begin{equation}
\label{eq:thefamousfour}
\Sigma(168)\,, \;\; \Sigma(72 \varphi)\,, \;\; \Sigma(216 \varphi) \,\;\; \text{and} 
\;\;\;  \Sigma(360 \varphi) \;.
\end{equation}
Essentially in this case the potentially dangerous invariants factorize $\I^{(4,4)} \to \I^{(2,2)}\I^{(2,2)}$ 
and the latter have the same invariants as the groups in Eq.~\eqref{eq:thefamousfour}.
\vspace{0.2cm}

Below we would like to reflect upon this rather general statement via examples and argue that even ``family irreducible'' type cases may be suitable in (many) perturbative type-models.
\begin{itemize}
\item \emph{$\Sigma(360\varphi)$ model independent:}
The most suitable candidate is $\Sigma(360\varphi)$ since the first new invariants appear only
at the $\I^{(4,4)}$ level c.f. Tab~\ref{tab:candidates}. 
The discussion of the previous subsection, c.f. Eq.~\eqref{eq:hm}, 
suggests that the most severe constraints could come from  
$s \to d$ transitions. 
We shall attempt at a rough estimate 
of the real part of $\Delta S = 2$ transitions. 
In the notation of Eq.~\eqref{eq:Leff} the effective Lagrangian assumes the 
following  form,
\begin{equation}
\label{eq:bsp}
\delta {\cal L}_{\rm eff}^{\Delta S =2} = \frac{\kappa^4}{\Lambda^2}   
\sum_n c_n \, \I_{n}(\Sigma(360 \varphi))^{abc2}_{1stu}  \left(  \bar d \, \tran_a^{\phantom{x} s} \tran_b^{\phantom{x} t} \tran_c^{\phantom{x} u}  \, s_L \right)    \left(  \bar d\, \tran_1^{\phantom{x} 2} \, s_L \right) \; ,
\end{equation}
where the transition matrices could be either $\tran_U$ or $\tran_D$ \eqref{eq:tran}.
The symbol $\kappa$ denotes 
the Yukawa expansion parameter \eqref{eq:discrete}.  It appears 
to the fourth power because of the four additional Yukawa matrices. 
We can now ask the following question: \emph{How small does $\kappa$ need to be in order for $C_{\rm dMFV}$ \eqref{eq:Leff} to satisfy the same
kind of experimental bounds as for $C_{\rm MFV}$  found in reference \cite{MFV}?}
The discussion of the  subsection, 
c.f. Eq.~\eqref{eq:hm}, suggests that 
$s \to d$ could be induced at first order in 
$\lambda$ as compared to  order
$|V_{ts} V_{td}^*| \sim \lambda^5$ in MFV.
The total $\Delta S = 2$ transition could therefore 
be ${\cal O}(\lambda^6)$ as compared to 
${\cal O}(\lambda^{10})$ in MFV.
According to our reflection above 
the condition is 
$\kappa^4 / \lambda^4 \simeq 1$ and
therefore  $\kappa \simeq \lambda \simeq 0.2$.

\begin{figure}[h]
 \centerline{\includegraphics[width=5.2in]{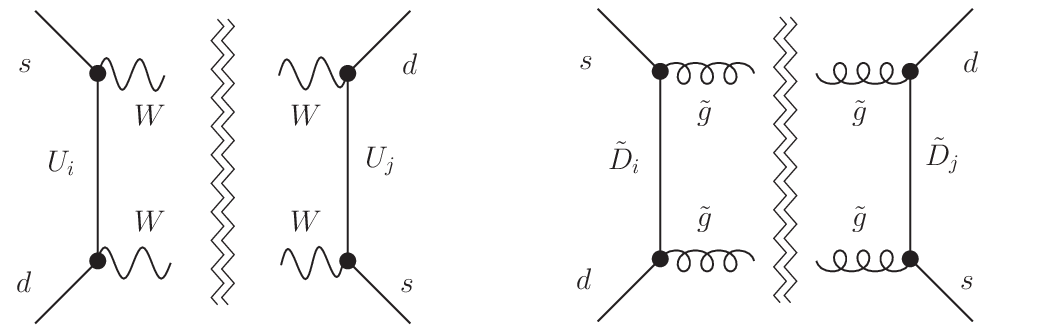}
 }
 \caption{\small \emph{``family reducible'':} 
 The double wiggly lines indicate cuts 
 through the diagrams where there is no (horizontal) 
 family charge flowing.
 (Left) SM box diagram.
  (Right) An example of a gluino contribution in the R-parity conserving MSSM.}
 \label{fig:examples_reducible}
 \end{figure}

\begin{figure}[h]
 \centerline{\includegraphics[width=5.2in]{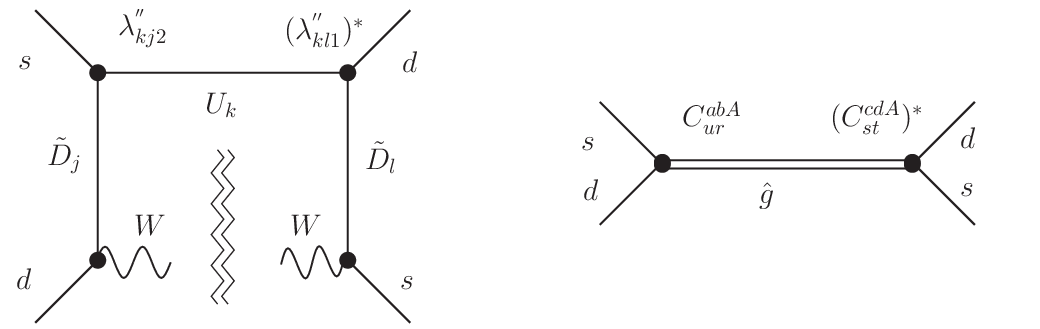}
 }
 \caption{\small \emph{``family irreducible'':} 
 The double wiggle lines indicate cuts 
 through the diagrams where there is no (horizontal) 
 family charge flowing.
 (Left) An example of a squark contribution in the R-parity violating MSSM e.g. \cite{Misiaketal}. 
Yet, this diagram does not generate new flavour structure in  discrete MFV.  (Right) Engineered  example which is family ireducible and does leads to new invariants in dicrete MFV. }
 \label{fig:examples_irreducible}
 \end{figure}

\item \emph{MSSM soft terms and $\Sigma(360\varphi)$:}
In the MSSM some additional flavour structure  enters through the soft terms,
e.g. the squark mass matrix $\tilde m_{ Q}^2$, which can be considered to be a ${\cal T}^{(1,1)}$ tensor. 
It was suggested a long time ago in the spirit of MFV \cite{hallrandall} 
that the nine  parameters of the hermitian $\tilde m_{ Q}^2$ could be organized into 
a Yukawa matrix  expansion:
\begin{equation}
\label{eq:soft}
(\tilde m_{ Q}^2)_r^{\phantom{x}c}  = \tilde m^2 \left(  a_1 \delta_r^{\phantom{x}c} + 
b_1^{(1)} (\tran_U)_r^{\phantom{x}c} + b_1^{(2)} (\tran_D)_r^{\phantom{x}c}  + 
b_2^{(1)} ( \tran_U \tran_D + {\rm h.c.})_r^{\phantom{x}c}  + ....
\right) \; .  \nonumber
\end{equation}
The first correction of the type given in Eq.~\eqref{eq:bsp}, in this expansion, would be given by
\begin{equation}
 (\delta \tilde m_{ Q}^2)_r^{\phantom{x}d}    =         b_3^{(1)} \, \I_{n}(\Sigma(360 \varphi))^{abcd}_{rstu} \,\tran_a^{\phantom{x} s} \tran_b^{\phantom{x} t}  \tran_c^{\phantom{x} u} \; .
\end{equation}
Note the assumption of the  the Yukawa hierarchy translates into  
${\cal O}( b_1) \simeq {\cal O}(  b_2 \kappa^{-2}) \simeq {\cal O}(  b_3 \kappa^{-4} )$ for the 
coefficients
\footnote{The assumption of a Yukawa hierarchy is not always imposed in the literature; e.g. \cite{noYH}. 
The finite dimension of the matrices makes the general series collapse at order $b_5$. Although the expansion 
contains more parameters than unknowns  predictivity results from 
the assumption that the $b_i$ are of the same order.
Moreover CKM and mass hierarchies also help in this respect.}.

\item \emph{family irreducible examples (c.f. Fig~\ref{fig:examples_irreducible}):} \\[0.1cm]
1. In the \emph{R-parity violating MSSM}, which is ``family irreducible'' Fig.~\ref{fig:examples_irreducible}(left), 
the fact that each vertex has to be 
$D_q$-invariant prevents the generation of 
non factorizable $\I^{(4,4)}$-invariants. 
This effectively happens because the R-parity 
violating vertices $\lambda^{''}_{ijk} U^i D^j D^k$ (U,D are superfields) 
do not allow for new structures because 
$\R{3} \times \R{3} \times \R{3}$ contains
the trivial representation only once for the groups in Eq.~\eqref{eq:thefamousfour}\footnote{To be even more concrete, in MFV the structure is given by $\lambda_{ijk}^{''\, \rm MFV} = 
a_1 \epsilon_{ajk} (\yuk{U} \yuk{D}^\dagger)^{\phantom{x} a}_{i} + a_2 \epsilon_{abc} 
(\yuk{U})^{\phantom{x} a}_{i} (\yuk{D})^{\phantom{x} b}_{j}(\yuk{D})^{\phantom{x} c}_{k} + \dots $ provided the U(1) structure does not forbid them from the start 
\cite{noYH}.}. \\[0.1cm]
2. Lacking a concrete example, let us imagine an effective theory with an interaction vertex 
${\cal L}_{\rm eff} \sim C^{abA}_{rs} \, \left( \bar D^r  \tran_a^{\phantom{x}s} D_b \right)  \,      \hat g_A $.
The variable $A$ is an index of a non-trivial representation $\R{A}$ appearing in 
$\Rb{3} \times \Rb{3} \times \R{3} \times \R{3} =
\R{A} + \dots $. The symbol $C^{abA}_{rs}$ denotes a generalized Clebsch-Gordan coefficient that makes the interaction ${\cal D}_{Q_L}$ invariant. 
The scalar particle $\hat g$ is supposed to be heavy and, when integrated out, leads to $\Delta S=2$ structure 
with a (new) non-factorizable $\I^{(4,4)}$-structure c.f. Fig.~\ref{fig:examples_irreducible}(right).

\end{itemize}

\section{Epilogue}
\label{sec:epilogue}

Before contemplating the scenario we shall briefly summarize our main results.
The reduced symmetry leads generally to further invariants and renders the mass-flavour basis transformation 
matrices observable, which can also be seen as a direct consequence
of the absence of the Goldstone bosons themselves.
Moreover non-SU(3) invariants do upset MFV hierarchies  
in a rather anarchic way,  c.f. subsection \ref{sec:mass}.   
In section \ref{sec:invariants} we have established, through an argument based on the absence
of a $\R{27}$ for discrete SU(3) subgroups, that there necessarily are new invariants for
$\I^{(4,4)}$. The latter enter generic $\Delta F = 2$ transitions \eqref{eq:generic2}. In cases where
the $\Delta F= 2$ process is generated via two $\Delta F = 1$ subprocesses, which we called ``family reducible'', 
the invariant factorizes: $\I^{(4,4)} \to \I^{(2,2)}\I^{(2,2)}$. For the latter the groups $\Sigma(168)$, $\Sigma(72 \varphi)$, $\Sigma(216 \varphi)$ and $\Sigma(360 \varphi)$ do provide enough symmetry to immitate 
SU(3) at this level 
and thus are valid candidates for a TeV-scale discrete MFV scenario with Yukawa expansion
parameters of the order of $\kappa_{\Sigma(360\varphi)} \simeq 0.2$. Models which are not of the ``family reducible''-type may
still be viable candidates; especially if they are perturbative.
An overview of the number of invariants
is given in Tab.~\ref{tab:candidates}.  Below we shall add a few not necessarily connected thoughts on MFV and
the framework proposed here.

\begin{itemize}

\item \emph{Origin of discrete symmetry:} 
One might of course ask the question about the origin of such 
discrete symmetries.
They might originate  from compactifications in String Theory, where it was found that 
the trihedral group $\Delta(54)$ can appear  \cite{ratz} or they could appear from the breaking of a continuous symmetry, c.f. \cite{Heidelberg} for a recent investigation.
The latter has to happen, presumably, at some high energy in order not to make the so far unseen familons too visible.  
We would like to add that whether global
symmetries originate from local ones or not can have subtle
physical  consequences \cite{KraussWilczek}.

\item \emph{This and that:} Surely it is possible that the groups ${\cal D}$ \eqref{eq:discrete} are of
different types. We have focused on  ${\cal D}_{Q_L}$, which gouverns the 
$D(U)_L \to D(U)_L'$ transitions.
The groups ${\cal D}_{D_R}$ would matter once we consider $D_R \to D_{R(L)}'$ type
operators, where again the groups 
\eqref{eq:thefamousfour} would provide
most protection. 
Needless to say that if the question of flavour is not
linked to a scale close to the TeV-scale,
and the breaking of  $D_q$ (or $G_q$) 
happens at high scale, then experimental 
bounds do not favour any particular groups.
Though in the MSSM for example 
the implementation of MFV is related to supersymmetry breaking through 
the soft terms \cite{hallrandall,MFV} \eqref{eq:soft} and this in turn 
suggests a link of flavour to the hierarchy problem.

\item \emph{Embeddings:}  The formulation \eqref{eq:discrete}  could be 
refined by constructing a discrete subgroups of 
$G_q$ \eqref{eq:Gq} which does not factor into 
direct products of SU(3) subgroups.
Much in the same way as the discrete subgroup $\Sigma(81) \subset U(3) \iso U(1) \times SU(3)\,$\footnote{$\Sigma(81) \supset \Delta(27)$ was first introduced in ref. \cite{Sigma(81)} and 
discussed in further detail in appendix B of ref. 
\cite{Claudia&Michael}.}
 cannot be written as a direct product of a
discrete SU(3) and U(1) subgroup; $\Sigma(81) \notin {\cal D}_{U(1)} 
\times {\cal D}_{SU(3)} $. One might wonder what the consequences for the invariants are.

\item \emph{Model of (d)MFV:} 
MFV is an empirically motivated
effective field theory approach. Up to now 
no explicit model of MFV has been constructed\footnote{Yet, in practice  anomaly mediated 
supersymmetry breaking, with CKM structure, 
is close to MFV \cite{MFV}.}. 
Though the seeds of a scenario were put forward in 
\cite{FMtop,FJM}. As hinted at in the introduction the relation 
between the MFV scale $\sqrt{16 \pi^2 C_{\rm MFV}}$ and the breaking scale(s)
$f_F$ can only be answered model by model.
It has to be added that a model of MFV without the input of the CKM 
and mass structure seems to be at the same level 
of difficulty as constructing a theory of flavour
which has proven to be a hard problem.
One should not forget that 
besides being predictive and testable MFV has 
other appealing properties:
In an R-parity violating MSSM, MFV  provides enough protection to evade bounds on the proton decay \cite{NoProtonDecay}. 
MFV also serves as a reference point for any model
with flavour structure and facilitates comparison of different models. 

\end{itemize}

Our aim, in this work, was to point out general issues of implementing 
MFV via a discrete group. We would hope that this work would be of some help
for further investigations towards more specific models, where for one reason or another
one or the other invariant does not turn out to be as menacing as in the essentially model
independent approach followed here. 


\section*{Acknowledgments}
RZ is particularly grateful to Gino Isidori for discussions which 
originated this work and to 
 Christoph Luhn, Thorsten Feldmann and Christopher Smith for very illuminating conversations. 
Further discussion and correspondence are acknowledged with Nick Evans, Jonathan Flynn, Claudia Hagedorn, Stephen King,  Sebastian J\"ager, Alexander Merle 
and many of the participants of the Kazmiercz Flavianet workshop 23-27 of July in 2009, where 
a large part of this work was first presented. We are grateful to Patrick Ludl for updating us on 
his great work on $(C)$- and $(D)$-groups.
 RZ gratefully acknowledges the support of an advanced STFC fellowship.

\appendix
\setcounter{equation}{0}
\renewcommand{\theequation}{A.\arabic{equation}}

\section{Examples of invariants in the flavour basis}
\label{app:examples}

In this appendix we discuss characteristic invariants of specific groups 
(in the flavour basis).
From section \ref{sec:necessity} we already know that new invariants are present at the level of the  effective theory. The aim of this appendix 
is to present a few instructive (concrete) examples.
We shall use the notation \eqref{eq:para3} for the left handed $D$-quarks.

\subsection{Crystal groups $\Sigma$}
\label{app:crystal}
We shall discuss $\Sigma(60)$ and $\Sigma(168)$ which are both instructive. 

\subsubsection*{The group $\Sigma(60)$}
The representations of $\Sigma(60) \iso  A_5 
\iso I_{\rm cosahedral} \subset SO(3)$ have for instance been
 studied in \cite{A5}. 
There are two real 3D irreps 
which we shall denote by $\R{3}_1$ and $\R{3}_2$. 
Their product representation takes the following form
\begin{equation}
\label{eq:S60_33}
\R{3}_i \times  \R{3}_i  =  \R{1} + \R{3}_i +  \R{5} \;, \qquad i = 1..2 \; .
\end{equation}
The $\R{3}_i$ on the right hand side (RHS) of \eqref{eq:S60_33}, with  \eqref{eq:para3} , 
 reads 
\begin{equation}
\label{eq:S60}
\R{3}_i  \sim (\bar D_3 D_2 - \bar D_2 D_3, \bar D_1 D_3 - \bar D_3 D_1,
\bar D_2 D_1 - \bar D_1 D_2   )^T \;.
\end{equation}
Since $\Rb{3}_i = \R{3}_i$  the singlet is obtained by simply taking the scalar 
product of the vector above 
\begin{equation}
\I^{(2,2)}_{ \Sigma(60)_2} =  \bar D_3 D_2 \bar D_3 D_2  + \dots  \; .
\end{equation}
The group $\Sigma(60)$ does allow for a $\Delta F = 2$ structure even in the absence of any Yukawa matrices.
The symmetry is simply not strong enough to constrain flavour transitions
in any way.

\subsubsection*{The group $\Sigma(168)$}

The isomorphisms of this group are: $\Sigma(168) \iso {\rm PSL}(2,\mathbb{F}_7) \iso 
{\rm GL}(3,\mathbb{F}_2)$ \cite{simple_discrete}. 
The irreps can be read off from \eqref{eq:168} and since 
the first non-trivial representations have the same
dimension this implies that they are identical \cite{simple_discrete},
$\Rb{3} \times \R{3} = \R{1} +  \R{8}$. The \R{8} is therefore real
but the difference appears at the level of product of two $\R{8}$ c.f. 
Eq.~\eqref{eq:88}.
As a consequence of the general discussion in subsection \ref{sec:necessity} the absence of a
$\R{27}$ implies further invariants. 
We may construct one of these invariants with the results given in \cite{simple_discrete} as follows: Consider the product
$\R{3} \times \R{3} = \Rb{3} + \R{6}$ and the information that the 6D is real, i.e. $\Rb{6} = 
\R{6}$, we may infer that the following product 
$\R{3} \times \R{3} \times \R{3} \times \R{3} = 1 \! \cdot \! \R{1} + ...$ contains the trivial representation once. The invariant
tensor may be read off from  \cite{simple_discrete} 
\begin{equation}
\label{eq:d-tensor}
d_{abcd} = d^{abcd} = \sum_{\alpha \beta} K_{ab}^\alpha K_{cd}^\beta C_{\alpha \beta} \; =  \;
 \left\{  \begin {array} {ll}  
 1/\sqrt{2}  & \quad abcd = \{1113,2221,3332\} \;\; \& \text{ cyclic}  \\
 0 & \quad \text{otherwise} 
\end{array}  \right.  \;,
\end{equation}
where cyclic refers to $1113 \to 3111 \to 1311 \to 1131$ etc. The first equality sign above is to be understood on the level of  indices and not at the level of tensors.
Note  that the invariant 
\begin{equation}
f = d^{abcd} D_a D_b D_c D_d \sim  D_3 (D_1)^3 + D_1 (D_2)^3 + D_2 (D_3)^3 \;,
\end{equation}
just corresponds to Klein's famous quartic invariant \cite{Klein}.
From \eqref{eq:d-tensor} we can build an invariant of the form 
\begin{equation}
\label{eq:crystal0}
(\I_{ \Sigma(168)_3})^{abcd}_{rstu} = d^{abcd} d_{rstu} \; .
\end{equation}
This invariant tensor, with field content \eqref{eq:generic2}, leads to terms of the form
\begin{equation}
\label{eq:crystal}
\I^{(4,4)}_{\Sigma(168)_3} \sim \bar D_1 \bar D_1 \tran_{1}^{\phantom{x}1} \tran_{2}^{\phantom{x}3} D_2 D_2  + \text{permutations} \;.
\end{equation}
Revealing  a rather anarchic structure of flavour transitions even in the 
flavour basis. Moreover the seventh tensor $\I^{(3,3)}_{\Sigma(168)_7}$, c.f.
Tab.~\ref{tab:candidates} can
easily be constructed
\begin{equation}
\label{eq:I3}
(\I^{(3,3)}_{\Sigma(168)_7})^{abc}_{rst} =  d_{abci} d^{irst}  \;.
\end{equation}
At last we would like to remark that the tensor $d_{ab}^{\phantom{xy}cd}=
  d_{abij} d^{ijcd}$  
\begin{equation}
\label{eq:Kronecker6}
d_{ab}^{\phantom{xy}cd}  \; =  \;
 \left\{  \begin {array} {ll}  
 1  & \quad a=b=c=d \\
 1/2 & \quad a=c , b=d \;\text{ or } \; a=d , b=c \\
 0 &  \quad \text{otherwise}
\end{array}  \right.  \;,
\end{equation}
acts, as expected, like a Kronecker symbol in the $\R{6}$-space.

\subsection{The trihedral groups $\Delta(3n^2)$ and $\Delta(6n^2)$}
\label{app:trihedral}

The main purpose of this subsection is to give some explicit non-SU(3) 
$\I^{(2,2)}$ invariants. In passing  we would like to mention
that as long as no real representations are generated, we have 
checked that for specific $n,k,l$ this is the case, all flavour transitions
are gouverned by the Yukawa matrices in the flavour basis. 
This fact is interesting but irrelevant to our work because the passage 
to the mass basis changes everything c.f. subsection \ref{sec:mass}.
At last we argue that the $D$-groups are subgroups of an 
appropriate $\Delta(6g^2)$ which is not known to the authors
from any other source.

\subsubsection*{The groups $\Delta(3n^2)$}
The group admits the following isomorphism \cite{BLW1}
$\Delta(3n^2)   \iso (\z{n} \times \z{n})  \rtimes \z{3}$.
As previously mentioned this group has only one and 3D irreps. 
They are labeled by the pair $(k,l)$ where 
$k,l = 0 .. n-1$ but $(k,l) \neq (0,0)$ (and additionally $(k,l)  \neq s(n/3,n/3)$ with $s = 1..2$ for $n = 3 \mathbb{Z}$)  and the following pairs, $(k,l) \iso (-k-l,,k) \iso (l,-k-l)$,  describe equivalent irreps.
The complex conjugate representation is obtained by reversing the sign of $k,l$, i.e. 
$ \Rb{3}_{(k,l)} = \R{3}_{(-k,-l)}$.
Anything relevant to us can be gained from the following Kronecker product 
\cite{BLW1,Delta3n2}
\begin{equation}
\label{eq:11133}
\Rb{3}_{(k,l)} \times \R{3}_{(k,l)} =  \R{3}_{(0,0)}   + \R{3}_{(-2k-l,k-l)} 
+ \R{3}_{(2k+l,l-k)}   \;,
\end{equation}
and the branching rules for  the RHS of \eqref{eq:11133} are
\begin{alignat}{2}
 \R{3}_{(0,0)}  &\to  \R{1}_{0,0}  +  \R{1}_{1,0}  +  \R{1}_{2,0}  \;, \qquad & &   
 \end{alignat}
and for $n = 3 \mathbb{Z}$ with  $k,l \in \mathbb{Z}n/3$, which reduces to $(k,l) = (0,\pm n/3)$ under equivalences,
\begin{alignat}{2}
 \R{3}_{(\frac{n}{3},\frac{n}{3})} 
 + \R{3}_{(-\frac{n}{3},-\frac{n}{3})}
 &\to  \R{1}_{0,1}  +  \R{1}_{1,1}  +  \R{1}_{2,1} +
  \R{1}_{0,2}  +  \R{1}_{1,2}  +  \R{1}_{2,2} \;,
\end{alignat}
 
there are nine one dimensional irreps.
With \eqref{eq:para3} and for $(k,l) = (0,n/3)$ they take on the form
\begin{eqnarray}
\label{eq:3n2_1}
\R{1}_{r,0} &\sim& \bar D^1 D_1  + \omega^{-r}  \bar D^2 D_2 + \omega^{r}  \bar D^3 D_3  \;, \nonumber \\
\R{1}_{r,1} &\sim& \bar D^2 D_1  + \omega^{-r}  \bar D^3 D_2 + \omega^{r}  \bar D^1 D_3  \;, \nonumber \\
\R{1}_{r,2} &\sim& \bar D^3 D_1  + \omega^{-r}  \bar D^1 D_2 + \omega^{r}  \bar D^2 D_3  \;,
\end{eqnarray}
and for $(k,l) = (0,-n/3)$  the roles of $\R{1}_{r,1/2}$ are reversed \cite{Delta3n2}.
Two of the generators, $a$ and $c$,  act in a non-trivial manner \cite{Delta3n2}:
 $a\circ  \R{1}_{r,s} = \omega^r \R{1}_{r,s}  \;,  c \circ  \R{1}_{r,s} = \omega^s \R{1}_{r,s}$
implying
$\R{1}_{r,s} \times \R{1}_{r',s'} = \R{1}_{r+r',s+s'} $. Note $\R{1}_{0,0}$ is therefore the only  singlet.  For any $n$ and $k,l$ there are at least five
invariant tensor at the level of $\I^{(2,2)}$, as compared to two for
 SU(3) c.f. Tab.~\ref{tab:candidates}.
For the symmetric contraction $\bar D^a\bar D^b D_c D_d$ 
the $\R{1}_{1,0} \times \R{1}_{2,0}$ invariant reads
\begin{alignat}{2}
\label{eq:3n2_I2}
  &     \I^{(2,2)}_{ \Delta(3n^2)_2 }  &=&  \;  \I^{(2,2)}_{ \Delta(3n^2)_1 }
- 3 (\bar D^1 D_1  \bar D^2 D_2 + \bar D^2 D_2  \bar D^3 D_3 + 
\bar D^1 D_1  \bar D^3 D_3) \;,
\nonumber \\ 
&  \I^{(2,2)}_{ \Delta(3n^2)_1 } &=& \; (\bar D^1 D_1 + \bar D^2 D_2 + \bar D^3 D_3)^2 \;,
\end{alignat}
where we have indicated the SU(3) invariant $\I_1$ for notational 
convenience. For the sake of completeness we shall indicate the explicit 3D  irreps, which 
can be obtained from appendix D of reference
\cite{Delta3n2}, up to a single transformation of the generator $a$, \begin{equation}
\label{eq:3n2_3}
\R{3}_{(-2k-l,k-l)} \sim ( \bar D_1 D_2 ,  \bar D_2 D_3 ,  \bar D_3 D_1)^T  \;, \quad
\Rb{3}_{(-2k-l,k-l)}  \sim ( \R{3}_{(2k+l,l-k)})^*  \; .
\end{equation}
From the explicit forms \eqref{eq:3n2_1} and \eqref{eq:3n2_3} it is
a simple matter to obtain the $\I^{(2,2)}$ invariants and even
beyond.
To this end we shall briefly discuss two cases of $\Delta(3n^2)$ which are popular in the literature. 
\begin{itemize}
\item[a)] $n =2: \Delta(12) \iso A_4$.
In fact this group was brought into particle physics as early as 1979 \cite{Wyler79}.
There is only one 3D irrep with $(k,l) = (0,1)$, which is real. 
The latter fact can either be checked explicitly, 
asserted from there being only one 3D irrep or inferred from the
fact that $A_4 \subset SO(3)$. The number of $\I^{(2,2)}$ invariants is seven and 
the reality of the $\R{3}$ allows to form an invariant
\begin{equation}
 \I^{(2,2)}_{ \Delta(3n^2)_4 }  = (\bar D_1 D_2)^2 + (\bar D_2 D_3)^2 
 + (\bar D_3 D_1)^2 
 \;, \text{and} \quad
\I_5  = \I_4 ^* .
\end{equation}
\item[b)] $n =3: \Delta(27)$ is the first group that admits nine one-dimensional 
irreps. The remaining ones completing the relation \eqref{eq:fund} is a complex conjugate pair of 3D irreps. There are nine $\I^{(2,2)}$ invariants, which can easily be obtained from \eqref{eq:3n2_1}.
\end{itemize}

\subsubsection*{The groups $\Delta(6n^2)$}
The group admits the following isomorphism \cite{BLW1}
$\Delta(6n^2)  \iso (\z{n} \times \z{n})  \rtimes S_3 \;.$
The irreps are  6, 3, 2 and 1D. 
The 6D representations $\R{6}_{(k,l)}$ are labeled by a pair $(k,l)$, where
$k = 0 .. (n \mi 1)$ and neither $k = 0$, $l=0$ nor $k + l = 0 \mod n$
(and additionally $(k,l)  \neq s(n/3,n/3)$ with $s = 1..2$ for $n = 3 \mathbb{Z}$).
Moreover the following six pairs $(k,l) \iso (-k-l,k) \iso (l,-k-l) \iso (-l,-k) \iso (k+l,-l)
\iso (-k,k+l) $, describe equivalent irreps. There are two types of 3D
irreps originating   from $\R{6}_{k,l}$ when 
$k+ l = 0 \mod n$ and  $(k,l) \neq (0,0)$. The two types of representations 
can therefore be labeled by $\Ri{3}{1}^{(l)}$ and  $\Ri{3}{2}^{(l)}$.
Complex conjugate irreps are obtained by reversing the sign of $(k,l)$ and $(l)$ respectively. For $n \in 3 \mathbf{Z}$ there are three further 
2D irreps denoted by $\Ri{2}{2}, \Ri{2}{3}, \Ri{2}{4}$ \cite{Delta6n2},
which are not relevant for $\I^{(2,2)}$ invariants. The  Kronecker product for the latter reads \cite{Delta6n2}:
\begin{equation}
\label{eq:3333}
(\Rib{3}{1}^{(l)} \times \Ri{3}{1}^{(l)}) 
\times (\Rib{3}{1}^{(l)} \times \Ri{3}{1}^{(l)}) =
(\Ri{3}{1}^{(0)} + \R{6}_{(l,l)}) \times (\Ri{3}{1}^{(0)} + \R{6}_{(l,l)}) \;,
 \end{equation}
where the explicit vectors on the RHS, using the parametrization \eqref{eq:para3}, are
\begin{eqnarray*}
\Ri{3}{1}^{(0)} = (\bar D^1 D_1, \bar D^2 D_2, \bar D^3 D_3)^T \; , \quad
\R{6}_{(l,l)}  = (\bar D^1 D_3, \bar D^3 D_2, \bar D^2 D_1, \bar D^1 D_2,\bar D^2 D_3, \bar D^3 D_1)^T  \;.
\end{eqnarray*}
Our form looks slightly more symmetric than the one
in reference \cite{Delta6n2} because we have chosen the $(l,l)$ rather than the $(-l,2l)$ representative.
The remaining relevant Kronecker products are:
\begin{alignat}{2}
\label{eq:33_6n2}
& \Ri{3}{1}^{(0)} \times \Ri{3}{1}^{(0)} &=& \;
\Ri{3}{1}^{(0)}   + \R{6}_{(0,0)} \nonumber  \;, \\
& \Ri{3}{1}^{(0)} \times  \R{6}_{(l,l)} &=& \;
3 \cdot \R{6}_{(l,l)}  \nonumber \;, \\
& \R{6}_{(l,l)}  \times  \R{6}_{(l,l)}  &=& \;
 \R{6}_{(0,0)} + \R{6}_{(2l,2l)}  + \R{6}_{(3l,0)} + \R{6}_{(0,3l)}
  + \R{6}_{(-l,2l)} +\R{6}_{(2l,-l)}   \; .
\end{alignat}
The RHS remains the same when   $\Ri{3}{1}$ is exchanged with
 $\Ri{3}{2}$ on the left hand side \cite{Delta6n2}. 
The relevant  branching rules are:
\begin{equation}
\Ri{3}{1}^{(0)} \to \R{1}_1 + \R{2}_1 \;, \quad \R{6}_{(0,0)} \to \R{1}_1 + \R{2}_1  
+  \R{1}_2 + \R{2}_2   \;.
\end{equation}

 The Clebsch-Gordan coefficient 
 of $ \Ri{3}{1}^{(0)}$  and  $\R{6}_{(0,0)}$ on the RHS of the top equation \eqref{eq:33_6n2} 
 are immediate from the ones of \eqref{eq:3333} and  the ones
 for $\R{6}_{(0,0)}$ are  \cite{Delta6n2} 
 $$\R{6}_{(0,0)} \sim (\bar D^1 D_6,\bar D^2 D_5,\bar D^3 D_4,\bar D^4 D_3,\bar D^5 D_2,\bar D^6 D_1)\;.$$ 
We have used  an obvious generalization of \eqref{eq:para3}.
It can be said that at the $\I^{(2,2)}$ level there are at least three invariants to be compared to two for  SU(3)  
c.f. Tab.~\ref{tab:candidates}. The Clebsch-Gordan coefficients allow
us to obtain them explicitly. 
We leave it to the reader to figure out the 
precise association of $(n,l)$ and the number of invariants.
The question is then how the singlets in Eq.~\eqref{eq:33_6n2} 
can be obtained from $\Ri{3}{1}^{(0)}$ and $\R{6}_{(0,0)}$. In both cases the generators 
$c$ and $d$ act trivially and then it remains to work out
which combination remains invariant under the
two remaining generators $a$ and $b$.  
Not surprisingly they are obtained by summing all the entries of the vectors. The correspondences are 
\begin{alignat}{2}
\label{eq:6n2_I}
&\Ri{3}{1}^{(0)}  \;\; \leftrightarrow  \;\; & & \I^{(2,2)}_{\Delta(6n^2)_1} 
 = \bar D^1 D_1 \bar D^1 D_1  + \bar D^2 D_2 \bar D^2 D_2 + \bar D^3 D_3 \bar D^3 D_3  \;,  \nonumber \\
&\R{6}_{(0,0)}  \;\; \leftrightarrow  \;\; & &  \I^{(2,2)}_{\Delta(6n^2)_2 } = \bar D^1 D_2  \bar D^2 D_1 + \bar D^1 D_3  \bar D^3 D_1 + 
\bar D^3 D_2  \bar D^2 D_3 \;.
\end{alignat}
It worth noting that both $\R{6}_{(0,0)}$ give rise to the same 
invariant $\I^{(2,2)}_2$ under the symmetric contraction 
$\bar D^a\bar D^b D_c D_d$. 
Note that $ \I_{\Delta(6n^2)_{1,2}}$ are different from $ \I_{\Delta(3n^2)_{1,2}}$ but since 
the two pairs are linearly dependent they are  effectively the same.

\subsubsection{The $(D)$-groups are subgroups of $\Delta(6n^2)$}
\label{app:Dgroups}
In the classic work of Miller et al \cite{Milleretal} the so-called 
$(C)$ and $(D)$-subgroups of SU(3) are defined as matrix groups. 
In~\cite{Ludl} it is shown that the $(C)$-groups are nothing but a special 
case of $\Delta(3n^2)$. We shall argue here that the $(D)$-groups are nothing
but subgroups of an appropriate $\Delta(6g^2)$.

In~\cite{Ludl},  the generators of the $(D)$-groups
have been worked out,
\begin{equation}
\label{eq:genD}
	H=\left(
	\begin{matrix}
	 \eta^a & 0 & 0 \\
	 0 & \eta^b & 0 \\
	 0 & 0 & \eta^{-a-b}
	\end{matrix}
	\right),\quad T=\left(\begin{matrix}
			 0 & 1 & 0 \\
			 0 & 0 & 1 \\
			 1 & 0 & 0
			\end{matrix}\right),\quad
	R=\left(
	\begin{matrix}
 	\delta^r & 0 & 0 \\
 	0 & 0 & \delta^s \\
 	0 & -\delta^{-r-s} & 0
	\end{matrix}
	\right),
\end{equation}
where
$\eta \equiv \exp( 2\pi i/n)$,  $\delta \equiv \exp( 2\pi i/d)$. They give rise to 
a collection of (not necessarily simple) six-parameter
subgroups $D(n,a,b;d,r,s)$ of $SU(3)$.
When viewed as a matrix
subgroup of~$SU(3)$ in its fundamental \R{3} representation, the
matrices belonging to these $(D)$-groups have exactly one non-zero entry
in every row and column. Furthermore, the non-zero entries are powers
of the g-th root of unity, with $g={\rm lcm}(n,d,2)$, where ${\rm lcm}$ stands for the lowest common multiple. The collection
of all such matrices evidently forms a group with $6\times g^2$
elements (the non-zero entries in the first and second row determine
the non-zero entry in the third, and there are six ways to place the
elements). As this group must be just the $\Delta(6n^2)$ group with
$n=g$, these $(D)$-groups are subgroups (proper or not) of the
$\Delta(6n^2)$ groups.  As such, they cannot possess an
irreducible representation whose dimension exceeds six.
An immediate consequence is that the group $D(n,a,b;d,r,s)$ shares 
the  two invariants Eq.~\eqref{eq:6n2_I} with 
$\Delta(6g^2)$.
The latter assertion can also be checked explicitly from the generators 
given in \eqref{eq:genD}. 

We would like to add that it was shown in \cite{Ludl2011} that whereas  the $(C)$-groups can  
be interpreted as irreducible representations of $\Delta(3g^2)$, 
this is not (always) the case for  $(D)$-groups with respect to  $\Delta(6g^2)$.

\section{Embedding of discrete groups into SU(3)}
\label{app:embedding}

We would like to settle the question of  whether it possible to approximate
an arbitrary SU(3) element (a basis transformation) by an element
of a discrete group ${\cal D} \subset$ SU(3)  suitably embedded 
into SU(3). The embedding of ${\cal D}$ into SU(3) ca be varied 
by conjugation with an arbitrary SU(3) matrix.
The problem therefore reduces to the question of whether there exist a
$D \in {\cal D}$ and $B \in SU(3)$ for a specific $A \in SU(3)$ 
such that
\begin{equation}
\label{eq:emb}
A \simeq B D B^\dagger \; .
\end{equation}
We shall argue below that for $\Delta(3n^2)$  this is possible. 
Eq.~\eqref{eq:emb} is true if $A$ and $D$ have (approximately) 
the same invariants.  The invariants are given by
the coefficients of the characteristic polynomial which are just the
trace of the matrix and the trace of the square of the matrix. A sufficient condition for the traces to be (approximately) the same  is that the eigenvalues are (approximately)  the same.
This immediately eliminates the crystal groups since there traces, i.e. 
characters, only assume very specific values. This can be inferred from the
character tables. 
We shall proceed our argument via the eigenvalues. An SU(3)
matrix has in general three eigenvalues of the form 
$\lambda_i = e^{i\phi_i}$ with the determinant condition $\phi_1 + \phi_2 + \phi_3 = 0 \mod 2 \pi$. The generators of $\Delta(3n^2)$
in the representation $\R{3}_{(k,l)}$ read \cite{Delta3n2}
\begin{equation}
a = \begin{pmatrix} 0 & 1 & 0  \\  0 & 0 & 1 \\ 1 & 0 & 0  \end{pmatrix},~~~~
c = \begin{pmatrix} \eta^l & 0 & 0  \\  0 & \eta^k & 0 \\ 0 & 0 & \eta^{-k-l}
\end{pmatrix},~~~~
d = \begin{pmatrix} \eta^{-k-l} & 0 & 0  \\  0 & \eta^l & 0 \\ 0 & 0 & \eta^k
\end{pmatrix}\  \,,
\end{equation}
with $\eta = \exp(2\pi i/n)$ and it is readily seen that the parameters
$k,l,n$ can be adjusted such that the eigenvalues, of for example $c$, 
are arbitrarily close to any pair of unitary complex numbers.
The third one is fixed in both cases 
by the determinant condition. For $\Delta(6n^2)$ this is also possible:
The elements $c_{3_1}^{a} d_{3_1}^{b}$, 
with generators $c_{3_1},d_{3_1}$ as given in \cite{Delta6n2}, 
approximate any two eigenvalues with
arbitrary precision for suitable 
$a,b ,n,l\in \mathbb{Z}$.

We conclude that $\Delta(3n^2)$ and $\Delta(6n^2)$ contrary to the crystal
groups can  be embedded into SU(3) such that one of its elements is arbitrarily close to any SU(3) element.

\end{document}